\documentclass[twocolumn,superscriptaddress,showpacs,amsmath,amssymb,aps,pra,floatfix]{revtex4-1}
\usepackage{graphicx}
\usepackage{bm}
\usepackage{amssymb}
\usepackage{amsmath}
\usepackage{amsthm}
\usepackage{amsfonts}
\usepackage{mathrsfs}
\usepackage{hyperref}

\begin{document}

\title{Gaussian-modulated coherent-state measurement-device-independent\\
 quantum key distribution}

\author{Xiang-Chun Ma}\affiliation{College of Science, National University of Defense Technology, Changsha 410073, People's Republic of China}\affiliation{Centre for Quantum Information and Quantum Control, Department of Physics \& Department of Electrical and Computer Engineering, University of Toronto, Toronto, Ontario, Canada M5S 3G4}

\author{Shi-Hai Sun}\affiliation{College of Science, National University of Defense Technology, Changsha 410073, People's Republic of China}

\author{Mu-Sheng Jiang}\affiliation{College of Science, National University of Defense Technology, Changsha 410073, People's Republic of China}

\author{Ming Gui}\affiliation{College of Science, National University of Defense Technology, Changsha 410073, People's Republic of China}

\author{Lin-Mei Liang}
\affiliation{College of Science, National University of Defense Technology, Changsha 410073, People's Republic of China}
\affiliation{State Key Laboratory of High Performance Computing, National University of Defense Technology, Changsha 410073, People's Republic of China}

\begin{abstract}
Measurement-device-independent quantum key distribution (MDI-QKD), leaving the detection procedure to the third partner and thus being immune to all detector side-channel attacks, is very promising for the construction of high-security quantum information networks. We propose a scheme to implement MDI-QKD, but with continuous variables instead of discrete ones, i.e., with the source of Gaussian-modulated coherent states, based on the principle of continuous-variable entanglement swapping. This protocol not only can be implemented with current telecom components but also has high key rates compared to its discrete counterpart; thus it will be highly compatible with quantum networks.
\end{abstract}

\maketitle

\section{Introduction}
Quantum key distribution (QKD), allowing a secret key between two legitimate parties (Alice and Bob) to be established \cite{Sca09}, has been applied to quantum information networks based on the trusted node or relay \cite{Fro13,Hug13}. However, in order to construct high-security and -performance networks, measurement-device-independent (MDI) QKD would be a very promising alternative since it not only removes all detector side-channel attacks, the most important security loophole of QKD implementations, by leaving the detection procedure to the untrusted relay but also supplies excellent performance with current technology \cite{Lo12,Bra12,Rub13,Liu13,Fer13,Tan13,Xu13}.

Measurement-device-independent QKD, which is a time-reversed Einstein-Podolsky-Rosen (EPR)-based QKD scheme \cite{Ina02}, consists of Alice and Bob respectively sending single-photon states to the third partner, Charlie, who makes a Bell-state measurement (BSM) and broadcasts his measurement results, and it has been a big step forward to bridge the gap between the theory and the real-world implementation of QKD \cite{Lo12}. Based on the idea of discrete-variable entanglement swapping and two-photon interference, the BSM can postselect the entanglement states between Alice and Bob and does not disclose the information about encodings, so this protocol allows the legitimate parties to establish the secure keys, which are independent of the measurement device, and all detection side-channel attacks are removed by leaving the detection procedure to the third partner.

As a counterpart, motivated by continuous-variable entanglement swapping \cite{Fur98,Bra98}, here, we propose a scheme to implement the MDI-QKD with continuous variables instead of discrete ones, i.e., with the source of Gaussian-modulated coherent states; thus we can confirm the security of this protocol by the optimality of Gaussian attacks \cite{Gar06,Nav06,Ren09L,Lev13L}. We show that with respect to this protocol two different reconciliation strategies, direct reconciliation (DR) and reverse reconciliation (RR), can be used to extract the secure keys even though this protocol seems symmetric for Alice and Bob compared to Charlie, and both are high performance. The detection side-channel attacks against continuous-variable QKD, such as the wavelength attack \cite{Ma13}, the calibration attack \cite{Jou13A}, the local oscillator (LO) intensity attack \cite{Ma13A}, and the saturation attack \cite{Qin13}, also have been excluded, and the expensive and low-detection-efficiency single-photon detectors used by discrete-variable MDI-QKD are also replaced by the lower-cost and higher-detection-efficiency balanced homodyne detectors (BHDs). Hence this protocol for continuous-variable MDI-QKD not only can be implemented with current technology but also has high key rates, like the conventional one-way continuous-variable (CV) QKD protocol [e.g., the Grosshans-Grangier protocol (GG02 protocol) \cite{Gro02,Gro03N}; see details in Ref.~\cite{Wee12R} and the references therein).

This paper is structured as follows. In Sec.~\ref{sec:Protocol}, we describe the protocol of continuous-variable MDI-QKD. In Sec.~\ref{sec:Estimation}, we give the security bounds of this protocol in DR and RR respectively against one-mode attack. In Sec.~\ref{sec:Discussion}, we extend the results of Sec.~\ref{sec:Estimation} to the asymmetric channel case and discuss the imperfect detections. Finally, Sec.~\ref{sec:Conclusion} is used for the conclusion of this paper.

\section{Protocol description}\label{sec:Protocol}
\begin{figure}[h]
 \includegraphics[width=.8\columnwidth]{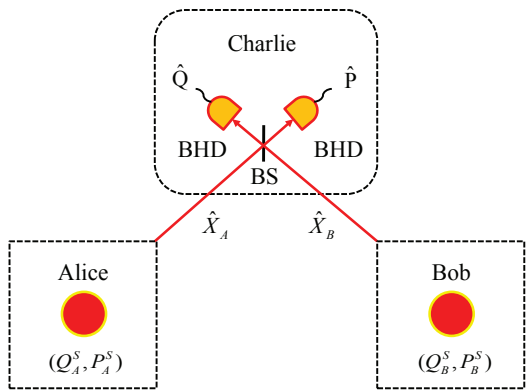}
 \caption{\label{fig:1}(Color online) Schematic setup of the continuous-variable MDI-QKD protocol. BHD: balanced homodyne detector; BS: balanced beam splitter.}
\end{figure}

This continuous-variable MDI-QKD, whose schematic setup is shown in Fig.~\ref{fig:1}, consists of the following four steps.

1. \textit{Preparation}. Alice and Bob each prepare coherent states in the phase space and send them to the third partner, Charlie, simultaneously, as shown in Fig.~\ref{fig:1}. Here the input modes can be described as $\hat{X}_{A/B}=X^S_{A/B}+\hat{X}^N_{A/B}$ for Alice and Bob, respectively, where $X^S_{A/B}$ are classical encoding variables with centered Gaussian distribution of zero mean and variance $V_S$ and $\hat{X}^N_{A/B}$ are vacuum modes. For all quadratures $\hat{Q}$ and $\hat{P}$ of coherent states, they are defined as $\hat{X}\in\{\hat{Q},\hat{P}\}$. The overall variance $V:=V(\hat{X}_A)$ of Alice's initial mode is given by $V=V_S+1$ in shot-noise units, where $V_S$ is the modulation variance mentioned before, and here we assume that Bob's variance of his mode is the same as Alice's, which they can agree on before key distribution without loss of generality \cite{Note}.

2. \textit{Measurement}. Charlie combines these two input modes with a balanced beam splitter (BS) and makes a continuous-variable BSM \cite{Fur98,Bra98} on the two modes with two BHDs shown in Fig.~\ref{fig:1}, i.e., one port detecting $\hat{Q}$ quadrature and the other $\hat{P}$ quadrature. Thus he will get the measurements, for example, $\hat{Q}_A-\hat{Q}_B$ and $\hat{P}_A+\hat{P}_B$ over the lossless and noiseless channels, up to a multiplier of $1/\sqrt{2}$ introduced by the BS \cite{Fur98,Bra98}. Then, he broadcasts these measurement results to Alice and Bob. Note that the LO used by Charlie is sent by either Alice or Bob, and before that, both Alice and Bob have defined the same signal-modulation reference frame by manipulating their respective LO beams (see details in Appendix \ref{sec:Define}).

3. \textit{Parameter estimation and security extraction}. Alice and Bob reveal part of their encodings, and based on Charlie's measurement results, they estimate the channel transmissions and excess noises. To establish the correlated data and secure keys, either Alice or Bob subtracts her or his encodings from Charlie's measurement results. For convenience, we assume that Bob implements this subtraction procedure since the protocol is symmetric; that is, Bob will take the data ($\hat{Q}+\sqrt{T_2}Q^S_B$) and $(\hat{P}-\sqrt{T_2}P^S_B)$, denoted as $\hat{Q}_{B^{'}}$ and $\hat{P}_{B^{'}}$, as estimations of Alice's encodings to establish the secure keys, where $T_2$ is the estimated channel transmission between Bob and Charlie. Since Eve does not know Bob's encodings, she does not know Alice's encodings accurately either from just the publication of quadratures $\hat{Q}$ and $\hat{P}$. Of course, she can learn part of the information from $\hat{Q}$ and $\hat{P}$.

4. \textit{Data postprocessing}. Alice and Bob extract the secret keys from their raw data using the current error correction and privacy amplification techniques \cite{Jou11} after they calculate the secret key rate between them.

\section{Estimation of security bounds}\label{sec:Estimation}
\begin{figure}[h]
 \includegraphics[width=.7\columnwidth]{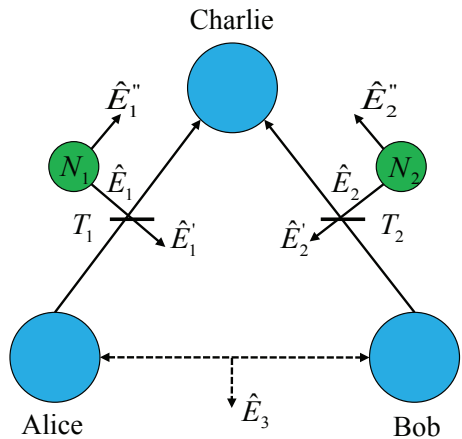}
 \caption{\label{fig:2}(Color online) Entangling cloner attack corresponding to two Markovian memoryless Gaussian channels with no interactions. Eve interacts Alice's and Bob's modes with her half of each EPR pair, respectively, and stores her ancillary modes $\hat{E}^{'}_1, \hat{E}^{''}_1, \hat{E}^{'}_2, \hat{E}^{''}_2$ in her quantum memory to acquire information by collective measurements at any time of the data-processing procedure. $N_1$ ($N_2$) denotes the variance of each mode of the first EPR pair (the second EPR pair). The mode $\hat{E}_3\in\{\hat{Q},\hat{P}\}$ is the virtual mode disclosed to Eve by Charlie's BSM and it can be taken as a classical variable.}
\end{figure}

To estimate the security bounds of our protocol, we consider the entangling cloner shown in Fig.~\ref{fig:2} to bound Eve's information. In the security analysis of conventional one-way CVQKD protocols, the collective Gaussian attacks up to an appropriate symmetrization of the protocols are considered to be the optimal general attacks \cite{Ren09L,Lev13L}. The entangling cloner is the most powerful and practical example of a collective Gaussian attack \cite{Pir08,Wee10} and is shown to be optimal for a single or one-mode channel \cite{Ma13A}. But in two-way protocols, the optimal attack is not clear with respect to two interaction channels, and the entangling cloner attack has only been demonstrated to be optimal in the hybrid two-way protocol \cite{Pir08}. In this work, we restrict our analysis to two Markovian memoryless Gaussian channels, which do not interact with each other and thus can be reduced to a one-mode channel \cite{Pir13N}. Hence, in this sense, the two independent entangling cloner attacks, one in each of the untrusted channels of our protocol, are reduced to one-mode attacks and thus can be taken as the optimal one-mode collective Gaussian attack (the most powerful attack corresponding to two memory Gaussian channels that interact with each other is analyzed in another work and is not analyzed in this paper; see the Note). This attack consists of Eve interacting on Alice's and Bob's modes with her half of each EPR pair, respectively, and the quantum channels are replaced by two beam splitters with transmissions $T_1$ and $T_2$, respectively. Then, she collects all the modes to store them in her quantum memory and makes collective measurements on these modes to acquire information at any time during the classical data-postprocessing procedure implemented by Alice and Bob.

As mentioned before, Bob's recast data are obtained by subtracting his own encodings from Charlie's publications. Before describing them, we first give the expressions for the BSM results $\hat{Q}$ and $\hat{P}$ under the entangling cloner attack shown in Fig.~\ref{fig:2}. They can be written as
\begin{equation}\label{eq:Q}
\begin{split}
\hat{Q}&\!=\!(\!\sqrt{T_1}\hat{Q}_A\!+\!\sqrt{1-T_1}\hat{Q}_{E_1}\!)\!-\!(\!\sqrt{T_2}\hat{Q}_B\!+\!\sqrt{1-T_2}\hat{Q}_{E_2}\!),\\
\hat{P}&\!=\!(\sqrt{T_1}\hat{P}_A\!+\!\sqrt{1-T_1}\hat{P}_{E_1})\!+\!(\sqrt{T_2}\hat{P}_B\!+\!\sqrt{1-T_2}\hat{P}_{E_2}),
\end{split}
\end{equation}
up to the multiplier of $1/\sqrt{2}$ mentioned before, which can be incorporated with the left-hand sides of the above equations. Here, $E_1$ and $E_2$ are Eve's EPR modes, whose variances are $N_1$ and $N_2$, respectively. $N_1$ and $N_2$ are used to simulate the variances of the practical channel excess noises $\varepsilon_A$ and $\varepsilon_B$, respectively; that is, $\varepsilon_A=(1-T_1)(N_1-1)/T_1$ for the channel between Alice and Charlie, $\varepsilon_B=(1-T_2)(N_2-1)/T_2$ for the channel between Bob and Charlie, and both are referred to their respective channel inputs. $T_1$ and $T_2$ are the respective channel transmissions of the channel between Alice or Bob and Charlie, and both can be estimated in the parameter-estimation procedure. Then, we can recast Bob's data as
\begin{equation}\label{eq:QB}
\begin{split}
\hat{Q}_{B^{'}}&\!=\!(\sqrt{T_1}\hat{Q}_A\!+\!\sqrt{1-T_1}\hat{Q}_{E_1})\!-\!(\sqrt{T_2}\hat{Q}^N_B\!+\!\sqrt{1-T_2}\hat{Q}_{E_2}),\\
\hat{P}_{B^{'}}&\!=\!(\sqrt{T_1}\hat{P}_A\!+\!\sqrt{1-T_1}\hat{P}_{E_1})\!+\!(\sqrt{T_2}\hat{P}^N_B\!+\!\sqrt{1-T_2}\hat{P}_{E_2}).
\end{split}
\end{equation}

Since the set of data in Eqs.~(\ref{eq:QB}) is a noisy version of Alice's encodings, restricted to one-mode attack, this protocol is equivalent to the conventional one-way CVQKD protocol with heterodyne detection \cite{Wee04}. In this sense, we can use the conventional standard methods to analyze its security bounds. Like for one-way CVQKD, two different strategies of reconciliation, DR and RR, can be used to extract secure keys. In DR, Alice's encodings are taken as the referential raw keys; thus Bob tries to guess them and reconcile his data to be identical to them by virtue of the additional classical side information sent by Alice. In RR Bob's recast data are taken as the raw keys; therefore Alice tries to make her encodings identical to them, requiring Bob to send side information. Note that Eve's entangling cloner attacks for these two reconciliation procedures are different. In DR, Eve just guesses Alice's encodings, and Bob's encodings have no contributions for her, so restricted to a one-mode attack, the entangling cloner attack on Bob's mode is no use to her except for reducing the mutual information between Alice and Bob. However, in RR, Eve tries to guess Bob's recast data, including not only Alice's encoding component but also the noise component, so she can acquire information with the help of entangling cloner attacks on both channels, as shown in Fig.~\ref{fig:2}.

Before calculating the secret key rates of this protocol in DR and RR, we first compute the Shannon information between Alice and Bob, and then in the following sections we bound Eve's information using the standard method (see \cite{Gar07,Pir08}) since this protocol is equivalent to the conventional one-way CVQKD protocol with heterodyne detection \cite{Wee04}.

Assuming the symmetry of both quadratures, the mutual information between Alice and Bob can be given by
\begin{equation}\label{eq:IAB}
I_{AB^{'}}=\log_2\frac{V_{B^{'}}}{V_{B^{'}|A}}.
\end{equation}
Note that they are identical in DR and RR and there is no multiplier of $1/2$ out the front since two quadratures are used to generate the secure keys, which is the same case as in conventional one-way CVQKD with heterodyne detection. The terms $V_{B^{'}}$ and $V_{B^{'}|A}$ are the variance and conditional variance of Bob's recast data $\hat{Q}_{B^{'}}$ and $\hat{P}_{B^{'}}$ in Eqs.~(\ref{eq:QB}). Since the terms on the right-hand sides of Eqs.~(\ref{eq:QB}) are mutually linearly independent, the variance $V_{B^{'}}:=\langle\hat{Q}_{B^{'}}^2\rangle=\langle\hat{P}_{B^{'}}^2\rangle$ ($\langle\hat{Q}_{B^{'}}\rangle=\langle\hat{P}_{B^{'}}\rangle=0$) is obtained by
\begin{equation}
V_{B^{'}}=T_1V+(1-T_1)N_1+T_2+(1-T_2)N_2:=b_v,
\end{equation}
and the conditional variance on Alice's encodings $X^S_A$ is given by
\begin{equation}
V_{B^{'}|A}=T_1+(1-T_1)N_1+T_2+(1-T_2)N_2:=b_0,
\end{equation}
using the formula of conditional variance defined as \cite{Poi94,Gra98}
\begin{equation}\label{eq:CVar}
V_{X|Y}=V(X)-\frac{|\left<XY\right>|^2}{V(Y)}.
\end{equation}
All the variances are in units of shot-noise level. Next, we calculate the secret key rates between Alice and Bob in DR and RR, respectively, by bounding Eve's information.

\subsection{Direct reconciliation}
In DR, Charlie's publication results will disclose some information, which is equal to giving Eve the virtual mode $\hat{E}_3$ shown in Fig.~\ref{fig:2}. Hence, Eve's information about Alice's encodings consists of the Shannon information $I_{AE_3}$ since $X_{E_3}$ ($\in\{Q,P\}$) is a classical variable and the Holevo information $\chi_{AE_A}$. The two kinds of information may partly contain each other, but we take the superposition as zero for simplicity. Therefore, the secret key rate can be given by
\begin{align}\label{eq:KDR}
K_{DR}=\beta I_{AB^{'}}-I_{AE_3}-\chi_{AE_A},
\end{align}
where $\beta$ is the efficiency of reconciliation. The Shannon information $I_{AB^{'}}$ is given by Eq.~(\ref{eq:IAB}). $I_{AE_3}$ bounds Eve's knowledge about Alice's encodings directly learned from Charlie's publication results $Q$ and $P$, and it can be taken as the classical information since Charlie's measurement for each pulse is individual (e.g., Alice and Bob can wait to send the next signal pulse until they receive the measurement result of the last pulse). The Holevo bound $\chi_{AE_A}$ describes Eve's information obtained from the entangling cloner shown in Fig.~\ref{fig:2}.

We first compute $I_{AE_3}$. Since Eve's modes $E^{''}_1$, $E^{''}_2$ can reduce the uncertainty of modes $E^{'}_1$ and $E^{'}_2$, respectively \cite{Gro03Q}, she can reduce the uncertainty of publication results $Q$ and $P$; i.e., the variance of mode $E_3\in\{Q,P\}$ conditioned on $E^{''}_1$, $E^{''}_2$ can be obtained by
\begin{equation}\label{}
V_{E_3|E^{''}_1,E^{''}_2}=T_1V+(1-T_1)/N_1+T_2V+(1-T_2)/N_2,
\end{equation}
using $V_{E_1|E^{''}_1}=1/N_1$ and $V_{E_2|E^{''}_2}=1/N_2$ \cite{Gro03Q}, and the conditional variance $V_{E_3|A,E^{''}_1,E^{''}_2}$ can also be given by
\begin{equation}\label{}
V_{E_3|A,E^{''}_1,E^{''}_2}=T_1+(1-T_1)/N_1+T_2V+(1-T_2)/N_2.
\end{equation}
So, assuming symmetry of both quadratures, the Shannon information $I_{AE_3}$ can be calculated as
\begin{equation}\label{eq:IAE}
I_{AE_3}=\log_2\frac{V_{E_3|E^{''}_1,E^{''}_2}}{V_{E_3|A,E^{''}_1,E^{''}_2}}.
\end{equation}

The Holevo information $\chi_{AE_A}$ can be written as
\begin{equation}\label{eq:XAE}
\chi_{AE_A}=S(E_A)-S(E_A|A),
\end{equation}
where $E_A$ denotes Eve's modes $E^{'}_1$, $E^{''}_1$, and $S(E_A)$ can be computed with the symplectic eigenvalues of the covariance matrix
\begin{equation}\label{}
\gamma_{E_A}(V,V)=\begin{pmatrix}
             e_{v1}\mathbb{I}&     \varphi_1\sigma_z\\
             \varphi_1\sigma_z&     N_1\mathbb{I}
             \end{pmatrix},
\end{equation}
where $\varphi_1=\sqrt{T_1(N_1^2-1)}$ and $e_{v1}=(1-T_1)V+T_1N_1$. Here $e_{v1}$ is the variance of mode $E^{'}_1$,  and the conditional variance on Alice's encodings is given by $e_0=(1-T_1)+T_1N_1$. $\mathbb{I}$ and $\sigma_z$ are Pauli matrices. The symplectic eigenvalues of this covariance matrix are given by
\begin{equation}
\lambda_{1,2}=\sqrt{\frac{\Delta\mp\sqrt{\Delta^2-4D}}{2}},
\end{equation}
where $\Delta=e^2_{v1}+N_1^2-2\varphi_1^2$ and $D=(e_{v1}N_1-\varphi_1^2)^2$. Hence, the von Neumann entropy of Eve's state is given by
\begin{equation}\label{eq:SE}
S(E_A)=G\left(\frac{\lambda_1-1}{2}\right)+G\left(\frac{\lambda_2-1}{2}\right),
\end{equation}
where $G(x)=(x+1)\log_2(x+1)-x\log_2x$. $S(E_A|A)$ can be obtained by the conditional covariance matrix
$\gamma_{E_A|A}=\gamma_{E_A}(1,1)$,
and its symplectic eigenvalues are given by
\begin{equation}
\lambda_{3,4}=\sqrt{\frac{A\mp\sqrt{A^2-4B}}{2}},
\end{equation}
where $A=e^2_0+N_1^2-2\varphi_1^2, B=(e_0N_1-\varphi_1^2)^2$. Thus, the conditional entropy is
\begin{equation}\label{eq:SEA}
S(E_A|A)=G\left(\frac{\lambda_3-1}{2}\right)+G\left(\frac{\lambda_4-1}{2}\right).
\end{equation}

\begin{figure}[h]
 \includegraphics[width=\columnwidth]{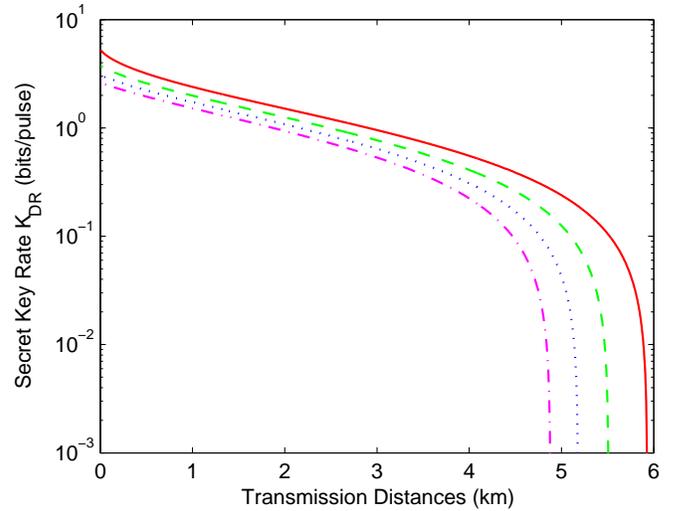}
 \caption{\label{fig:3}(Color online) Secret key rates vs transmission distances between Alice, Charlie, and Bob for DR with symmetric channels. From top to bottom, excess noise is selected as 0, 0.005, 0.01, and 0.015, which are typical values in experiments \cite{Jou13N}. Alice and Bob's modulation variance is set to be optimal, and the reconciliation efficiency is 0.95, which is an appropriate value (see \cite{Jou11}). Here, fiber loss is 0.2 dB/km.}
\end{figure}

With Eqs.~(\ref{eq:IAE}) and (\ref{eq:XAE}), we can bound Eve's information for DR and then compute the secret key rate $K_{DR}$ in Eq.~(\ref{eq:KDR}). We plot it in Fig.~\ref{fig:3} for a symmetric channel case where $T_1=T_2$ and excess noises in each channel are also identical.

From Fig.~\ref{fig:3}, we can see that this protocol is very sensitive to channel loss and excess noise in DR for the symmetric channel case, and transmission distances are limited to 15 km (3-dB limit), the same as for the one-way CVQKD protocol with DR, or shorter due to the fact that Bob's data contain some modulation vacuum noise, which is detrimental to him, and Charlie's BSM discloses some information to Eve.

\subsection{Reverse reconciliation}\label{sec:Reverse}
In RR, the secret key rate can be written as
\begin{align}\label{eq:KRR}
K_{RR}=\beta I_{AB^{'}}-I_{B^{'}E_3}-\chi_{B^{'}E},
\end{align}
where $I_{AB^{'}}$ is also given by Eq.~(\ref{eq:IAB}), $I_{B^{'}E_3}$ describes the information disclosed by Charlie, and $\chi_{B^{'}E}$ quantifies Eve's Holevo information about Bob's recast data ($\hat{Q}_{B^{'}}$ or $\hat{P}_{B^{'}}$) by entangling cloner attack. The latter two quantities are calculated as follows.

The information $I_{B^{'}E_3}$ about Bob's recast data $\hat{Q}_{B^{'}}$ and $\hat{P}_{B^{'}}$ disclosed by Charlie's BSM can be written as
\begin{equation}\label{eq:IBE}
I_{B^{'}E_3}=\log_2\frac{V_{E_3}}{V_{E_3|B^{'}}}.
\end{equation}
where there is no factor of $\frac{1}{2}$, as the previous section mentioned. To compute the variance of mode $E_3\in\{Q,P\}$ and the conditional variance $V_{E_3|B^{'}}$, we recast the measurement quadrature $\hat{Q}$ ($\hat{P}$) in Eqs.~(\ref{eq:Q}) as $\hat{Q}=\hat{Q}_{B^{'}}-\sqrt{T_2}Q_B$ ($\hat{P}=\hat{P}_{B^{'}}+\sqrt{T_2}P_B$), so these variances can be given, respectively, by
\begin{equation}
\begin{split}
V_{E_3}&=\langle(\hat{Q})^2\rangle=\langle(\hat{P})^2\rangle\\
&=T_1V+(1-T_1)N_1+T_2V+(1-T_2)N_2,
\end{split}
\end{equation}
\begin{equation}
V_{E_3|B^{'}}=T_2V_S=T_2(V-1).
\end{equation}
Then, using the above equations, Eq.~(\ref{eq:IBE}) can be obtained.

The Holevo information $\chi_{B^{'}E}$ can be obtained by
\begin{equation}\label{eq:XBE}
\chi_{B^{'}E}=S(E)-S(E|B^{'}),
\end{equation}
where $E$ denotes Eve's modes $E^{'}_1$, $E^{''}_1$, $E^{'}_2$, $E^{''}_2$. $S(E)$ can be computed with the symplectic eigenvalues of the covariance matrix,
\begin{equation}\label{}
\gamma_{E}=\begin{pmatrix}
             e_{v1}\mathbb{I}&     \varphi_1\sigma_z&  0& 0\\
             \varphi_1\sigma_z&     N_1\mathbb{I}&    0& 0\\
             0& 0&   e_{{v2}}\mathbb{I}&     \varphi_2\sigma_z\\
             0& 0&   \varphi_2\sigma_z&     N_2\mathbb{I}
\end{pmatrix}_{8\times8},
\end{equation}
where $e_{v2}=(1-T_2)V+T_2N_2$ and $\varphi_2=\sqrt{T_2(N_2^2-1)}$. This covariance matrix can be written as $\gamma_{E_A}\bigoplus\gamma_{E_B}$, so $S(E)=S(E_A)+S(E_B)$, where $S(E_A)$ is given by Eq.~(\ref{eq:SE}) and $S(E_B)$ is obtained by replacing $T_1$ and $N_1$ with $T_2$ and $N_2$ in Eq.~(\ref{eq:SE}). Likewise, $S(E|B^{'})$ can be calculated by symplectic eigenvalues of the conditional covariance matrix $\gamma^{Q_{B^{'}},P_{B^{'}}}_E$, which can be obtained by \cite{Gar07}
\begin{equation}
\gamma^{Q_{B^{'}},P_{B^{'}}}_E=\gamma_E-\sigma_{EB^{'}}(\textbf{X}\gamma_{B^{'}}\textbf{X})^{MP}\sigma^T_{EB^{'}},
\end{equation}
where
\begin{equation}
\begin{split}
\sigma_{EB^{'}}&=\begin{pmatrix}
\langle\hat{Q}_{E^{'}_1}\hat{Q}_{B^{'}}\rangle& 0\\
0& \langle\hat{P}_{E^{'}_1}\hat{P}_{B^{'}}\rangle\\
\langle\hat{Q}_{E^{''}_1}\hat{Q}_{B^{'}}\rangle& 0\\
0& \langle\hat{P}_{E^{''}_1}\hat{P}_{B^{'}}\rangle \\
\langle\hat{Q}_{E^{'}_2}\hat{Q}_{B^{'}}\rangle& 0\\
0& \langle\hat{P}_{E^{'}_2}\hat{P}_{B^{'}}\rangle\\
\langle\hat{Q}_{E^{''}_2}\hat{Q}_{B^{'}}\rangle& 0\\
0& \langle\hat{P}_{E^{''}_2}\hat{P}_{B^{'}}\rangle
\end{pmatrix}=
\begin{pmatrix}
\xi_1\mathbb{I}\\
\phi_1\sigma_z\\
-\xi_2\sigma_z\\
-\phi_2\mathbb{I}
\end{pmatrix},
\end{split}
\end{equation}
with
\begin{equation}
\begin{split}
\xi_1&=\sqrt{T_1(1-T_1)}(N_1-V),\\
\phi_1&=\sqrt{(1-T_1)(N_1^2-1)},\\
\xi_2&=\sqrt{T_2(1-T_2)}(N_2-1),\\
\phi_2&=\sqrt{(1-T_2)(N_2^2-1)},
\end{split}
\end{equation}
and $\gamma_{B^{'}}=\bigl(\begin{smallmatrix}bv & 0 \\
                                           0 & bv \end{smallmatrix}\bigr)$, $\textbf{X}=\bigl(\begin{smallmatrix}1 & 0 \\
                                           0& 1 \end{smallmatrix}\bigr)$. $MP$ stands for the Moore-Penrose inverse of a matrix.
For a straightforward calculation, the conditional covariance matrix can be recast as
\begin{equation}\label{}
\begin{split}
&\gamma^{Q_{B^{'}},P_{B^{'}}}_{E}=\\
&\begin{pmatrix}
             (e_{v1}-\frac{\xi^2_1}{b_v})\mathbb{I}&     (\varphi_1\!-\!\frac{\xi_1\phi_1}{b_v})\sigma_z&  \frac{\xi_1\xi_2}{b_v}\sigma_z&            \frac{\xi_1\phi_2}{b_v}\mathbb{I}\\
             (\varphi_1\!-\!\frac{\xi_1\phi_1}{b_v})\sigma_z&     (N_1-\frac{\phi_1^2}{b_v})\mathbb{I}&    \frac{\xi_2\phi_1}{b_v}\mathbb{I}&            \frac{\phi_1\phi_2}{b_v}\sigma_z\\
              \frac{\xi_1\xi_2}{b_v}\sigma_z&     \frac{\xi_2\phi_1}{b_v}\mathbb{I}&  (e_{v2}-\frac{\xi^2_2}{b_v})\mathbb{I}&     (\varphi_2\!-\!\frac{\xi_2\phi_2}{b_v})\sigma_z \\
             \frac{\xi_1\phi_2}{b_v}\mathbb{I}&      \frac{\phi_1\phi_2}{b_v}\sigma_z&   (\varphi_2\!-\!\frac{\xi_2\phi_2}{b_v})\sigma_z&     (N_2-\frac{\phi_2^2}{b_v})\mathbb{I}
             \end{pmatrix}.
\end{split}
\end{equation}
Calculating the symplectic eigenvalues of a four-mode covariance matrix is very challenging, and the standard method is as follows \cite{Gar07,Ser06}: first, we denote the four symplectic eigenvalues as $\nu_1$, $\nu_2$, $\nu_3$ and $\nu_4$, which satisfy
\begin{equation}\label{eq:Delta}
\begin{split}
&\Delta^4_1=\nu^2_1+\nu^2_2+\nu^2_3+\nu^2_4,\\
&\Delta^4_2=\nu^2_1\nu^2_2+\nu^2_1\nu^2_3+\nu^2_1\nu^2_4+\nu^2_2\nu^2_3+\nu^2_2\nu^2_4+\nu^2_3\nu^2_4,\\
&\Delta^4_3=\nu^2_1\nu^2_2\nu^2_3+\nu^2_1\nu^2_2\nu^2_4+\nu^2_1\nu^2_3\nu^2_4+\nu^2_2\nu^2_3\nu^2_4,\\
&\Delta^4_4=\nu^2_1\nu^2_2\nu^2_3\nu^2_4,
\end{split}
\end{equation}
where $\Delta^4_j$ ($j=1,2,3,4$) is the $jth$-order principal minor of $\gamma^{Q_{B^{'}},P_{B^{'}}}_E$, which is defined as the sum of the determinants of all the $2j\times2j$ submatrices of the $n\times n$ covariance matrix obtained by deleting $n-2j$ rows and the corresponding $n-2j$ columns \cite{Ser06}. Second, after calculating the principal minors of the conditional covariance matrix, we can solve Eqs.~(\ref{eq:Delta}) to get the symplectic eigenvalues of the four-mode conditional covariance matrix. However, it is very difficult. But, we can compute $S(E|B^{'})$ asymptotically. Note that $G(\frac{\nu-1}{2})\rightarrow\log_2\frac{e\nu}{2}+O(\nu^{-1})$ for $\nu\gg1$ \cite{Pir08}. This means that for a large variance $V$ of Alice's and Bob's modulated modes and $T\neq0,1$, we can use the above formula to compute the Holevo information. For a large modulation variance, the asymptotic eigenvalues of $\gamma^{Q_{B^{'}},P_{B^{'}}}_{E}$ can be given by
\begin{equation}\label{eq:eigen}
\nu_1=N_1, \nu_2=N_2,
\nu^2_3\nu^2_4=\frac{\Delta^4_4}{\nu^2_1\nu^2_2}.
\end{equation}
Then, $S(E|B^{'})$ can be obtained by
\begin{equation}
S(E|B^{'})=G\left(\frac{N_1-1}{2}\right)+G\left(\frac{N_2-1}{2}\right)+\log_2\frac{e^2\nu_3\nu_4}{4}.
\end{equation}
Then, the Holevo information $\chi_{B^{'}E}$ in Eq.~(\ref{eq:XBE}) can be attained with above equations.

\begin{figure}
 \includegraphics[width=\columnwidth]{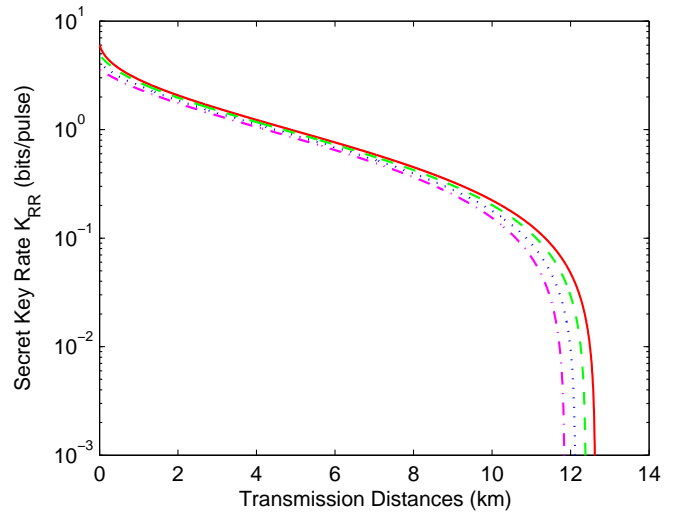}
 \caption{\label{fig:4}(Color online) Secret key rates vs transmission distances between Alice, Charlie, and Bob for RR with symmetric channels. From top to bottom, excess noise is selected as 0, 0.005, 0.01, and 0.015. Alice and Bob's modulation variance is also set to be optimal. Fiber loss is 0.2 dB/km, and the reconciliation efficiency is 0.95.}
\end{figure}

Hence, we can obtain the secret key rate in RR in Eq.~(\ref{eq:KRR}) by substituting Eqs.~(\ref{eq:IAB}), (\ref{eq:IBE}), and (\ref{eq:XBE}) into it. We plot the secure key rate $K_{RR}$ as a function of transmission distances between Alice, Charlie, and Bob in Fig.~\ref{fig:4} for the symmetric channel case. The performance in RR is higher than that in DR, which is analogous to the case of conventional one-way CVQKD, but in conventional one-way CVQKD the RR protocol has no loss limit when the channel excess noise is zero. Since the vacuum noise of Bob's mode reduces the mutual information between Alice and Bob and Charlie's BSM discloses some information, the transmission distances are also very limited, with typical experimental parameters used in current CVQKD implementations \cite{Jou13N}, except for the modulation variance of Alice and Bob, which is set to be optimal. Note that the calculation of the Holevo information $\chi_{B^{'}E}$ in Eq.~(\ref{eq:XBE}) is for the case of large modulation variance of Alice and Bob. However, in Appendix \ref{sec:Asymptotic}, we show that even for infinitely strong modulation and perfect reconciliation efficiency the transmission distances are still short, and also the improvement is limited.

\section{Discussion}\label{sec:Discussion}
\begin{figure}
 \includegraphics[width=\columnwidth]{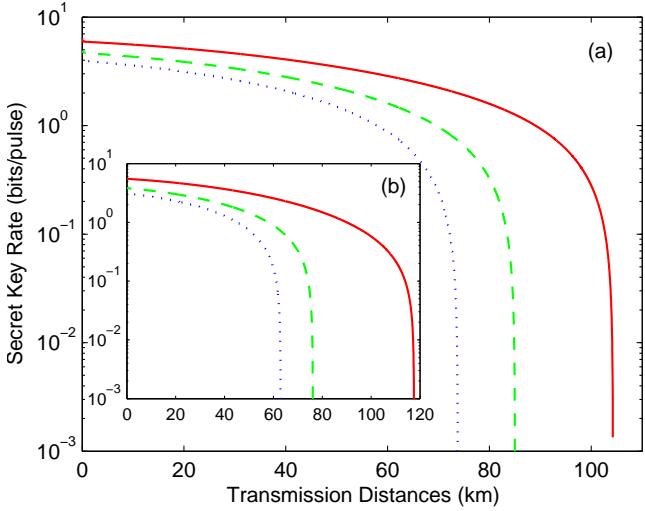}
 \caption{\label{fig:5}(Color online) Secret key rates vs transmission distances between Alice, Charlie, and Bob for (a) RR and (b) DR in the asymmetric channel case with the optimal modulation variance. Excess noise are selected as 0 (solid lines), 0.005 (dashed lines), and 0.01 (dotted lines), for both channels referred to their respective channel transmissions $T_1$ and $T_2$. Both BSM relays are close to Alice's station, and the distance is set to be 10 m. Fiber loss is 0.2 dB/km, and the reconciliation efficiency is 0.95.}
\end{figure}

As shown in the previous sections, we set Bob to recast his data by subtracting his encodings from Charlie's BSM results; however, if Alice recasts her data as Bob does and Bob keeps his encodings, the same results as above will be obtained. Although the performance is not very good for symmetric channels, we show that this protocol will exhibit high performance for asymmetric channels ($T_1\neq T_2$), as shown in Fig.~\ref{fig:5}.

In Fig.~\ref{fig:5}, we set Charlie's BSM relay close to Alice's station, for example, 10 m, and find that both DR and RR have excellent performances over long distances between Alice, Charlie, and Bob with experimental realistic conditions. Moreover, if we set Charlie's BSM relay close to Bob's station, the cases are a little more complicated due to the respective channel excess noises. However, in this setting, this protocol is very close to the conventional one-way CVQKD protocol except Charlie's BSM discloses part of the information to Eve. Therefore, in this sense, for this protocol DR has 3-dB limit, and RR has no loss limit if there is no channel excess noise and $T_2\rightarrow 1$. We do not show these results in the figure. Of course, we can reverse the above cases in Fig.~\ref{fig:5} to get high performance with Charlie's BSM relay close to Bob's station just by having Alice recast her data if channels have excess noise.

Finally, we point out that the imperfections of Charlies's homodyne detections, such as detection efficiency and electronic noise, can be included in the channel transmission and excess noise, respectively; thus we can not necessarily consider the imperfections of Charlie's detections when computing the secure key rates. However, these imperfections will reduce the performance of this protocol rapidly if the BHD has low detection efficiency and high electronic noise. Hence using highly efficient BHDs in BSM is necessary to improve the performance of this continuous-variable MDI-QKD.

\section{Conclusion}\label{sec:Conclusion}
In conclusion, we proposed a scheme to realize the idea of MDI-QKD, with a source of Gaussian-modulated coherent states. We showed that this protocol has higher performance in RR against a one-mode optimal attack than DR for the symmetric channel case, but both are limited to short distances; however, for asymmetric channels both have excellent performances and can be extended to current distances realized by the conventional one-way CVQKD. Moreover, the protocol almost exploits each pulse to generate keys and thus has high key rates compared to the discrete-variable MDI-QKD. Actually, this protocol has no basis choice or comparison, and each pulse except the ones used for parameter estimation contributes to the establishment of secure keys. In addition, the source can be easily generated with coherent light, and the whole protocol can be implemented experimentally with current technology, although the LO interference will be a little complicated. We hope to seek other methods to solve the problem of pulse synchronization and the reference frame calibration in future research.


\begin{acknowledgments}
The authors thank H.-K. Lo and his group for enlightening discussions and comments. This work is supported by the National Natural Science Foundation of China, Grants No. 61072071 and No. 11304391. L.-M.L. is supported by the Program for New Century Excellent Talents. X.-C.M. is supported by the Hunan Provincial Innovation Foundation for Postgraduates. X.-C.M. and M.-S.J. acknowledge support from NUDT under Grant No. kxk130201.
\end{acknowledgments}

\appendix
\section{Defining the reference frame}\label{sec:Define}
\begin{figure}[h]
 \includegraphics[width=.9\columnwidth]{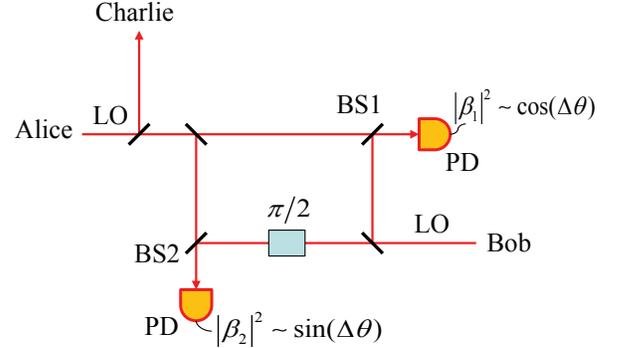}
 \caption{\label{fig:6}(Color online) Schematic setup for measuring phase difference between two classical LO beams. PD: photodetector; BS1, BS2: beam splitters; all beam splitters in the figure are balanced, or 50:50.}
\end{figure}

In this appendix, we discuss how to synchronize the pulses and define the reference frame between Alice, Bob, and Charlie by the manipulation of LO. The basic idea is that, if we can measure the phase difference of two LO beams sent by Alice and Bob, respectively, we can add this phase difference in one party's modulation of his or her signal beam, and thus the two signal modulations of Alice and Bob are implemented in the same reference frame. Since LO is a strong classical beam, we can combine two LO beams in a balanced beam splitter so that they interfere with each other; then we can measure one port's interference output to get the phase difference of the two LO beams. The schematic setup is shown in Fig.~\ref{fig:6}.

In this protocol, we have Alice send the LO beam to Charlie, who then splits it into two beams with a balanced beam splitter for his two balanced homodyne detectors and uses them to measure the quadratures $\hat{Q}$ and $\hat{P}$. First, Alice splits her LO beam into two beams, one sent to Charlie and the other to Bob. Then, Bob splits the received LO beam from Alice and his own LO beam into two beams, respectively, and combines them with BS1 and BS2 so they interfere with each other, as shown in Fig.~\ref{fig:6}. Then, we use each photodetector on one port of both BS1 and BS2, to detect the intensity of interfered beams. Note that, we add a $\pi/2$ phase on Bob's one split LO beam in order to accurately measure the phase difference. We denote the amplitude of Alice's LO beam that interferes with Bob's as $\alpha e^{i\theta_A}$ and denote Bob's LO beam as $\alpha e^{i\theta_B}$, provided that both of them have identical intensities. Relative to the LO beams, Alice's and Bob's classical signal beams are phase modulated into $\alpha^A_S e^{i(\theta_A+\phi_A)}$ and $\alpha^B_S e^{i(\theta_B+\phi_B)}$, respectively, before attenuating the quantum level. $\alpha^A_S$ and $\alpha^B_S$ are their respective signal beam intensities, and $\phi_A$, $\phi_B$ are modulated phases. Then, when two LO beams interfere with BS1, the amplitude of one port can be written as
\begin{equation}
\beta_1=\frac{\alpha e^{i\theta_A}+\alpha e^{i\theta_B}}{\sqrt{2}}=\sqrt{2}\alpha e^{\frac{i(\theta_A+\theta_B)}{2}}\cos \left(\frac{\theta_A-\theta_B}{2}\right).
\end{equation}
The PD output of BS1 is obtained by
\begin{equation}\label{eq:Beta1}
|\beta_1|^2=2|\alpha|^2\cos^2\left(\frac{\theta_A-\theta_B}{2}\right)\!=\!|\alpha|^2[1+\cos(\theta_A-\theta_B)].
\end{equation}
Likewise, the PD output of BS2 can be attained as
\begin{equation}\label{eq:Beta2}
|\beta_2|^2\!=\!|\alpha|^2\{1+\cos[\theta_A-(\theta_B+\pi/2)]\}=|\alpha|^2[1+\sin(\theta_A-\theta_B)].
\end{equation}
With Eqs.~(\ref{eq:Beta1}) and (\ref{eq:Beta2}), we can accurately compute the phase difference $\Delta\theta:=\theta_A-\theta_B$ of Alice's and Bob's LO beams. When Bob modulates his signal beam, he adds the phase difference $\Delta\theta$ and the initial phase $\phi_B$ together as the modulated phase. Thus, the amplitude of Bob's signal beam can be written as $\alpha^B_S e^{i(\theta_B+\phi_B+\Delta\theta)}=\alpha^B_S e^{i(\theta_A+\phi_B)}$ , which has been defined in the same reference frame as Alice's.

However, realizing the above strategy experimentally may be complicated, and we just give a simple theoretical method and demonstrate the possibility of implementing this whole protocol of continuous-variable MDI-QKD. Other strategies to solve the problem of pulse synchronization and the reference frame calibration might exist. We note that in Refs. \cite{Zha00,Zha02,Jia04}, homodyne detectors and LO beams are not needed to make the BSM in continuous-variable entanglement swapping; however, we are not sure whether their method of BSM is suitable for this protocol. Finally, we point out that, to relieve the emitters' burden and preserve the symmetry of Alice and Bob, this procedure of synchronization and the reference frame calibration can be implemented by Charlie without affecting the security of this protocol. That means Alice and Bob both send their LO beams to Charlie, who measures the phase difference of the two beams and then adds it to the signal beam of either Alice or Bob by modulation.

\section{Asymptotic key rate for RR with infinitely strong modulation}\label{sec:Asymptotic}
\begin{figure}
 \includegraphics[width=\columnwidth]{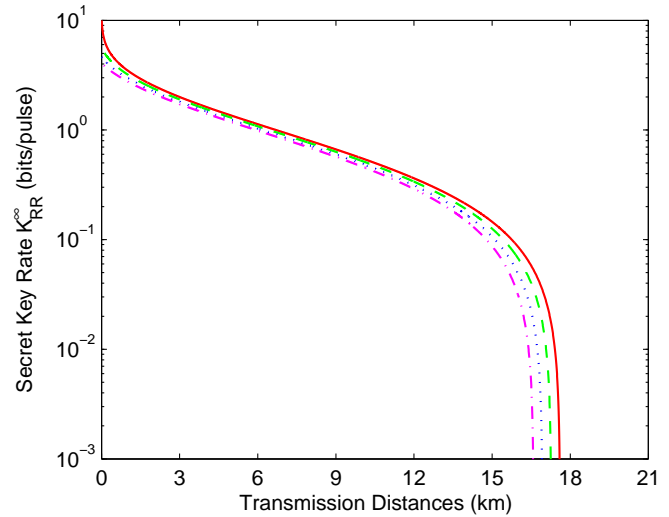}
 \caption{\label{fig:7}(Color online) Asymptotic key rates vs transmission distances between Alice, Charlie, and Bob for RR in the symmetric channel case with infinitely large modulation variance and perfect reconciliation efficiency, i.e., $V\rightarrow\infty$ and $\beta=1$. From top to bottom, excess noise is selected as 0, 0.005, 0.01, and 0.015. Fiber loss is 0.2 dB/km.}
\end{figure}

In Sec. \ref{sec:Reverse}, we obtained the Holevo information $\chi_{B^{'}E}$ in Eq.~(\ref{eq:XBE}) for RR for the case of large modulation variance of Alice and Bob. Using appropriate experimental parameters, e.g., a finite reconciliation efficiency $\beta=0.95$, and optimizing the modulation variance, we show that the transmission distances in the symmetric channel case for RR are also limited like in DR due to the vacuum noise of Bob's mode, which is different from the conventional one-way CVQKD as mentioned before. However, in this appendix, we point out that even for infinitely strong modulation and perfect reconciliation efficiency, the transmission distances are still limited, as shown in Fig.~\ref{fig:7}.

Figure \ref{fig:7} depicts the asymptotic key rate with infinitely large modulation variance for RR in the symmetric channel case, and we can see that the improvement in the achievable key rate is limited with respect to the modulation variance of Alice and Bob. In addition, we can easily check that the maximum transmission distance in the asymptotic case is extended by only about 2 km compared to the one in the case with $V=40$ and $\beta=1$. This means that the asymptotic calculation of the eigenvalues and Holevo information $\chi_{B^{'}E}$ in Eqs.~(\ref{eq:eigen}) and (\ref{eq:XBE}), respectively, is also applicable to the case of experimental realization with an appropriately large modulation variance. Finally, we point out that the asymptotic key rates for other cases in DR and RR with symmetric or asymmetric channels can also be easily obtained using the above method, i.e., by setting $V\rightarrow\infty$ and $\beta=1$ in the calculation for the key rates. For the purpose of experimental realization, i.e., using the experimentally realistic parameters, we do not give their results here.
%


\begin{thebibliography}{99}

\bibitem{Sca09} V. Scarani, H. Bechmann-Pasquinucci, N. J. Cerf, M. Du\v{s}ek, N. L\"{u}tkenhaus, and M. Peev, Rev. Mod. Phys. \textbf{81}, 1301 (2009).

\bibitem{Fro13} B. Fr\"{o}hlich, J. F. Dynes, M. Lucamarini, A. W. Sharpe1, Z. Yuan and A. J. Shields, Nature \textbf{501}, 69 (2013).

\bibitem{Hug13} R. J. Hughes, J. E. Nordholt, K. P. McCabe, R. T. Newell, C. G. Peterson and R. D. Somma, arXiv:1305.0305v2 [quant-ph] (2013).

\bibitem{Lo12} H. K. Lo, M. Curty, and B. Qi, Phys. Rev. Lett. \textbf{108}, 130503 (2012).

\bibitem{Bra12} S. L. Braunstein and S. Pirandola, Phys. Rev. Lett. \textbf{108}, 130502 (2012).

\bibitem{Rub13} A. Rubenok, J. A. Slater, P. Chan, I. Lucio-Martinez, and W. Tittel, Phys. Rev. Lett. \textbf{111}, 130501 (2013).
\bibitem{Liu13} Y. Liu \textit{et al.}, Phys. Rev. Lett. \textbf{111}, 130502 (2013).

\bibitem{Fer13} T. Ferreira da Silva, D. Vitoreti, G. B. Xavier, G. C. do Amaral, G. P. Temporao, and J. P. von der Weid, Phys. Rev. A \textbf{88}, 052303 (2013).

\bibitem{Tan13} Z. Tang, Z. Liao, F. Xu, B. Qi, L. Qian, and H.-K. Lo, arXiv:1306.6134 (2013).

\bibitem{Xu13} F. Xu, B. Qi, Z. Liao, and H.-K. Lo, Appl. Phys. Lett. \textbf{103}, 061101 (2013).

\bibitem{Ina02} H. Inamori, Algorithmica \textbf{34}, 340 (2002).

\bibitem{Fur98} A. Furusawa, J. L. S{\o}rensen, S. L. Braunstein, C. A. Fuchs, H. J. Kimble, and E. S. PolzikScience, \textbf{282}, 706 (1998).

\bibitem{Bra98} S. L. Braunstein and H. J. Kimble, Phys. Rev. Lett. \textbf{80}, 869 (1998).

\bibitem{Gar06} R. Garc{\'{\i}}a-Patr{\'{o}}n and N. J. Cerf, Phys. Rev. Lett. \textbf{97}, 190503 (2006).

\bibitem{Nav06} M. Navascu\'{e}s, F. Grosshans, and A. Ac\'{\i}n, Phys. Rev. Lett. \textbf{97}, 190502 (2006).

\bibitem{Ren09L} R. Renner and J. I. Cirac, Phys. Rev. Lett. \textbf{102} 110504 (2009).

\bibitem{Lev13L} A. Leverrier, R. Garc{\'{\i}}a-Patr{\'{o}}n, R. Renner, and N. J. Cerf, Phys. Rev. Lett. \textbf{110}, 030502 (2013).

\bibitem{Ma13} X.-C. Ma, S.-H. Sun, M.-S. Jiang, and L.-M. Liang, Phys. Rev. A \textbf{87}, 052309 (2013); J.-Z. Huang, C. Weedbrook, Z.-Q. Yin, S. Wang, H.-W. Li, W. Chen, G.-C. Guo, and Z.-F. Han, Phys. Rev. A \textbf{87}, 062329 (2013); J.-Z. Huang, S. Kunz-Jacques, P. Jouguet, C. Weedbrook, Z.-Q. Yin, S. Wang, W. Chen, G.-C. Guo, and Z.-F. Han, Phys. Rev. A \textbf{89}, 032304 (2014).

\bibitem{Jou13A} P. Jouguet, S. Kunz-Jacques, and E. Diamanti, Phys. Rev. A \textbf{87}, 062313 (2013).

\bibitem{Ma13A} X.-C. Ma, S.-H. Sun, M.-S. Jiang, and L.-M. Liang, Phys. Rev. A \textbf{88}, 022339 (2013).

\bibitem{Qin13} H. Qin R. Kumar and R. All\'{e}aume, Proc. SPIE \textbf{8899}, 88990N (2013).

\bibitem{Gro02} F. Grosshans and P. Grangier, Phys. Rev. Lett. \textbf{88}, 057902 (2002).

\bibitem{Gro03N} F. Grosshans, G. V. Asschee, J. Wenger, R. Brouri, N. Cerf, and P. Grangier, Nature (London) \textbf{421}, 238 (2003).

\bibitem{Wee12R} C. Weedbrook, S. Pirandola, R. Garc{\'{\i}}a-Patr{\'{o}}n, N. J. Cerf, T. C. Ralph, J. H. Shapiro, and S. Lloyd, Rev. Mod. Phys. \textbf{84}, 621 (2012).

\bibitem{Note} Note that, in this paper, we do not consider the imperfections of modulation which may cause the difference in modulation variances between Alice and Bob. The most general case is that Alice and Bob have different modulation variances instead of the same variance, which will be analyzed in future research.

\bibitem{Jou11} P. Jouguet, S. Kunz-Jacques, and A. Leverrier, Phys. Rev. A \textbf{84}, 062317 (2011).

\bibitem{Wee10} C. Weedbrook, S. Pirandola, S. Lloyd, and T. C. Ralph, Phys. Rev. Lett. \textbf{105}, 110501 (2010).

\bibitem{Pir08} S. Pirandola, S. Mancini, S. Lloyd, and S. L. Braunstein, Nat. Phys. \textbf{4}, 726 (2008).

\bibitem{Pir13N} S. Pirandola, New J. Phys. \textbf{15}, 113046 (2013).

\bibitem{Wee04} C. Weedbrook, A. M. Lance, W. P. Bowen, T. Symul, T. C. Ralph, and P. K. Lam, Phys. Rev. Lett. \textbf{93}, 170504 (2004).

\bibitem{Gar07} R. Garc{\'{\i}}a-Patr{\'{o}}n, Ph.D. thesis, Universit${\rm \acute{e}}$ Libre de Bruxelles, 2007.

\bibitem{Poi94} J. P. Poizat, J. F. Roch, and P. Grangier, Ann. Phys. (Paris) \textbf{19}, 265 (1994).

\bibitem{Gra98} P. Grangier, J. A. Levenson, and J. P. Poizat, Nature (London) \textbf{396}, 537 (1998).

\bibitem{Gro03Q} F. Grosshans, N. J. Cerf, J. Wenger, R. Tualle-Brouri, and P. Grangier, Quantum Inf. Comput.\textbf{ 3}, 535 (2003).

\bibitem{Jou13N} P. Jouguet, S. Kunz-Jacques, A. Leverrier, P. Grangier, and E. Diamanti, Nat. Photonics \textbf{7}, 378 (2013).

\bibitem{Ser06} A. Serafini, Phys. Rev. Lett. \textbf{96}, 110402 (2006).

\bibitem{Zha00} J. Zhang and K. Peng, Phys. Rev. A \textbf{62}, 064302 (2000).

\bibitem{Zha02} J. Zhang, C. Xie, and K. Peng, Phys. Lett. A \textbf{299}, 427 (2002).

\bibitem{Jia04} X. Jia, X. Su, Q. Pan, J. Gao, C. Xie, and K. Peng, Phys. Rev. Lett. \textbf{93}, 250503 (2004).



\end{thebibliography}
\end{document}